\documentclass[conference,english]{IEEEtran}
\IEEEoverridecommandlockouts

\usepackage[T1]{fontenc}
\usepackage{graphicx}
\usepackage{xcolor}
\usepackage[normalem]{ulem}
\usepackage{amsmath}
\usepackage{amsthm}
\usepackage{enumerate}
\usepackage{bm,bbm}
\usepackage{comment}
\usepackage{babel,csquotes}
\usepackage{microtype}
\usepackage{mathabx}
\usepackage{mathtools}

\usepackage[style=ieee, doi=false, isbn=false, url=false]{biblatex}
\renewcommand{\baselinestretch}{0.98}

\newtheorem{theorem}{Theorem}[section]

\newtheorem{lemma}[theorem]{Lemma}
\newtheorem{corollary}[theorem]{Corollary}

\input{macros.tex}

\DeclareMathOperator{\diag}{diag}

\newcommand{\keywords}[1]
{
  \small
  \textbf{\textit{Keywords---}} #1
}

\begin{document}

\title{Small-Noise Sensitivity Analysis of Locating Pulses in the Presence of Adversarial Perturbation}

\author{\IEEEauthorblockN{Meghna Kalra and Kiryung Lee \thanks{MK and KL were supported by NSF CAREER Award CCF-1943201.}}
\IEEEauthorblockA{\textit{Department of Electrical and Computer Engineering} \\
\textit{The Ohio State University}\\
Columbus, OH, USA}
\and
\IEEEauthorblockN{Maxime Ferreira Da Costa \thanks{MF is with the Laboratory of Signals and Systems (L2S), CNRS. His work was supported by ANR: ANR-20-IDEES-0002.}}
\IEEEauthorblockA{\textit{CentraleSup\'elec}\\
\textit{Universit\'e Paris-Saclay} \\
Gif-sur-Yvette, France }
}

\maketitle

\begin{abstract}
A fundamental small-noise sensitivity analysis of spike localization in the presence of adversarial perturbations and an arbitrary point spread function (PSF) is presented.
The analysis leverages the local Lipschitz property of the inverse map from measurement noise to parameter estimate.
In the small noise regime, the local Lipschitz constant converges to the spectral norm of the noiseless Jacobian of the inverse map. An interpretable upper bound in terms of the minimum separation of spikes, norms, and flatness of the PSF and its derivative, as well as the distribution of spike amplitudes is provided.
Numerical experiments highlighting the relevance of the theoretical bound as a proxy to the local Lipschitz constant and its dependence on the key attributes of the problem are presented.
\end{abstract}


\IEEEpeerreviewmaketitle
\section{Introduction}
\label{sec:intro}

The localization of spikes convolved with a known point spread function (PSF) arises in various applications, including imaging, array processing, radar, and communications. While this problem has a long history in signal processing~\cite{kay1981spectrum}, recent progress in super-resolution established theory and computational methods for the spike localization against or even below the Rayleigh resolution limit (\emph{e.g.}~\cite{schiebinger2018superresolution,diederichs2019well,chi2020harnessing,li2020approximate,li2021stable}).
Yet, most of the literature assumes the point spread function (PSF) is a Dirac distribution, or the noise is stochastic, typically modeled as Gaussian.
However, there are many practical applications in which the target pulse signal is convolved with a non-Dirac PSF and is corrupted with a structured or an adversarial noise. This yields a gap between the classical theoretical guarantees on super-resolution and their applicability in real-world use cases.
Given a fixed noise energy, structured noise can deteriorate the estimation performance more than random noise.
For example, the spike injection framework~\cite{li2023channel} considers spurious adversarial spikes consuming the entire noise energy. The error estimate on the parameters under this noise model has been shown to be related to the distance between the true and the faked signal components and can yield a significantly greater error estimate on the true signal parameter than random perturbation. This found interest in the context of physical layer security, where an agent transmitted over a wireless channel seeks to conceal her physical location from an eavesdropper. When the spike injection pattern is known to the receiver, the physical location can be preserved without degrading the decoding of legitimate parties.
In surface electromyography (EMG) modality~\cite{de2012inter}, the signal is given as the convolution of spikes representing neural activations and patient-dependent PSF, known as motor unit action potential (MUAP).
In this application, structural noise arises in the form of cross-talk, constituting neural signals from neighboring muscles
and introducing additional peaks in the signal reading.
This peak can be misinterpreted as the activation of the muscle under observation and can cause the EMG acquisition algorithm to make a wrong inference.

\vspace{1mm}
\noindent\textbf{Contributions:}
This paper establishes a fundamental analysis of the localization of spikes in the presence of non-Dirac PSF and adversarial noise. We are particularly interested in characterizing the phase transition between two regimes where stable estimation of the spikes is feasible and unfeasible.
We will quantify the phase transition condition given in an interpretable form depending on the minimum separation of the spikes, PSF characteristics, and amplitude dynamic range.
Our approach to the sensitivity analysis is based on the local Lipschitz property of the inverse map that maps an arbitrary noisy instance to a least squares estimator.
This local Lipschitz property describes how smoothly a function behaves in the vicinity of any point within its domain. Essentially, it implies that within any small neighborhood around a point, the change in the function's output does not exceed a certain rate when compared to the change in the input.
This framework enables the worst-case analysis in terms of the noise pattern.
To the best of the authors' knowledge, we have not found prior work on the fundamental analysis of spike localization in this worst-case scenario.

We focus on an asymptotic scenario of a small noise where the signal-to-noise ratio is sufficiently high. In this regime, the Lipschitz constant of the inverse map converges to the spectral norm of its noiseless Jacobian, which we use as a proxy to characterize the sensitivity of the model.
We start by providing an analytical expression of the noiseless Jacobian of the inverse map. In order to better highlight its dependency on the key attribute of the problem, such as the minimum separation between the pulses, the shape of the PSF, and the dynamic range of the amplitudes, we derive in Theorem~\ref{THM:MAIN} a novel simple and interpretable upper estimate of its spectral norm.
To achieve this goal, we rely on the recent advances characterizing the spectrum of weighted Vandermonde matrices~\cite{moitra2015super,AUBEL2019} through the analytical properties of the extremal solution of generalized Beurling--Selberg type approximation problems~\cite{vaaler1985ExtremalFunctions, carneiro2011SurveyBeurlingSelberg, carneiro2014ExtremalFunctions, vaaler2023NumberLattice}.

The fundamental sensitivity analysis in this paper contrasts with the classical Cramér-Rao bound in the following perspectives.  While the Cramér-Rao bound applies only to the class of unbiased estimators, most of the practical estimators for spike localization are biased.
Furthermore, except for a few examples, including the additive white Gaussian model, the computation of the Cramér-Rao bound is not easy.
On the contrary, our fundamental analysis provides a unique characterization in the worst-case noise model and applies regardless of the noise distribution, even to adversarial noise cases.

\vspace{1mm}
\noindent\textbf{Mathematical Notations:}
Vectors and matrices are written in small boldface $\bm{a}$ and capital boldface $\bm{A}$, respectively. $\bm{A}^*$ denotes the conjugate transpose of matrix $\bm{A}$. The notations $\norm{\bm{A}}$, $\norm{\bm{A}}_{\mathrm{F}}$, $\norm{\bm{A}}_{\infty \rightarrow 2}$ refer to the spectral norm, Frobenius norm, and the largest $\ell_2$ norm of the columns of a matrix $\bm{A}$. The maximum eigenvalue and singular value of $\bm{A}$ are denoted $\lambda_{\max}(\bm{A})$ or $\sigma_{\max}(\bm{A})$, respectively.
The shorthand notation $[n]$ denotes the set of integers $\{1,\ldots,n\}$. The number $\mathrm{j}$ is the basis of imaginary numbers, $\mathrm{j}^2 = -1$. For $T > 0$, the $K$-dimensional torus is written $\mathbb{T}^K \sim \left(\mathbb{R} / T \mathbb{Z} \right)^K$.

\vspace{1mm}
\noindent\textbf{Organization:}
Section \ref{sec:problem statement} lays the formulation of the problem statement and the signal model. Section \ref{sec:sensiti analysis} presents the main sensitivity analysis and provides the supporting lemmas and the main theorem. Section \ref{sec:proof} proposes a sketch of the proof of the main results. Section \ref{sec:numerical} presents the validation for the theoretical analysis. Finally, we conclude with Section \ref{conclsion}.

\section{Problem Statement}
\label{sec:problem statement}
Consider the parameter estimation of pulses from multiple snapshots.
The observed signal in the $\ell$th snapshot is written as
\begin{equation*}
y_\ell(t) = \sum_{k=1}^K x_{k,\ell}  g(t-\tau_k), \quad \ell \in [L] := \{1,\dots,L\}.
\end{equation*}
The signal consists of a superposition of $K$ pulses of known PSF $g$ centered at common locations $\{\tau_k\}_{k=1}^K \subset \mathbb{T}$ with varying amplitudes across snapshots.
The goal of the problem is to estimate the unknown locations in $\bm{\tau} = [\tau_1,\dots,\tau_K]^\mathsf{T}$ from the noisy Fourier transform coefficients of $y_\ell$ sampled at frequencies in $\{f_i = \frac{2i-N-1}{2T} : i \in [N]\}$ across $L$ snapshots.
Define $\bm{\gamma} \mapsto \bm{\Phi}_{\bm{\gamma}} \in \mathbb{C}^{N \times K}$ by $[\bm{\Phi}_{\bm{\gamma}}]_{i,k} = e^{-\mathsf{j} 2\pi \gamma_k f_i}$ where $\bm{\gamma} = [\gamma_1,\dots,\gamma_K]^\mathsf{T} \in \mathbb{T}^K$.
Let $\bm{Y} \in \mathbb{C}^{N \times L}$ collect all measurements so that $[\bm{Y}]_{i,\ell}$ denotes the Fourier transform $\widehat{y_\ell}(f) = \int_{-\infty}^\infty y_\ell(t) e^{-\mathsf{j} 2\pi f t} dt$ at frequency $f_i$, corrupted with additive noise, for $i \in [N]$ and $\ell \in [L]$.
Then $\bm{Y}$ is compactly written as
\begin{equation*}
    \bm{Y} = \bm{G} \bm{\Phi}_{\bm{\tau}} \bm{X} + \bm{Z}
\end{equation*}
where $\bm{G} = \diag\left(\widehat{g}(f_1), \dots, \widehat{g}(f_N) \right)$ is a diagonal matrix, where $\widehat{g}$ denotes the Fourier transform of $g$, $\bm X \in \mathbb{C}^{K \times L}$ satisfies $[\bm X]_{k,\ell} = x_{k,\ell}$, and $\bm{Z} \in \mathbb{C}^{N \times L}$ denotes additive noise. We assume that $\bm{G}$ has at least $K$ nonzero diagonal entries. Then $\bm{G} \bm{\Phi}_{\bm{\tau}}$ has full column rank.

We consider the least squares estimator given by
\begin{equation*}
\mathop{\mathrm{minimize}}_{\bm{\gamma}\in \mathbb{T}^K, \bm{\Upsilon} \in \mathbb{C}^{K \times L}} ~ \norm{\bm{Y}-\bm{G} \bm{\Phi}_{\bm{\gamma}} \bm{\Upsilon}}_{\mathrm{F}}^2
\end{equation*}
Given $\bm{\gamma}$, the optimal $\bm{\Upsilon}$ is given by $(\bm{G} \bm{\Phi}_{\bm{\gamma}})^\dagger \bm{Y}$.
Therefore, the same optimal estimator of $\bm{\tau}$ is obtained by
\begin{equation}
\label{eq:varpro}
\mathop{\mathrm{minimize}}_{\bm{\gamma}\in \mathbb{T}^K} ~ \norm{\bm{P}_{\bm{\gamma}}^\perp \bm{Y}}_{\mathrm{F}}^2
\end{equation}
where $\bm{P}_{\bm{\gamma}}^\perp := \bm{I}_N - \bm{G} \bm{\Phi}_{\bm{\gamma}} (\bm{G} \bm{\Phi}_{\bm{\gamma}})^\dagger$ denotes the projection onto the orthogonal complement of the column space of $\bm{G} \bm{\Phi}_{\bm{\gamma}}$.
Let $\hat{\bm{\tau}}$ denote a minimizer to \eqref{eq:varpro}.
In the noise-free scenario ($\bm{Z} = \bm{0})$ under the sufficient condition that $N > 2K$, the estimate is uniquely determined as the ground-truth $\bm{\tau}$.
The objective is to study the perturbation of the least-squares estimator as a function of worst-case additive noise $\bm{Z}$.

\section{Local Sensitivity Analysis}
\label{sec:sensiti analysis}

For the purpose of analysis, we assume up to a scaling of the pulse location and the PSF that $T=1$. We present a fundamental sensitivity analysis that is based on the local Lipschitz property of the inverse map $\bm{\psi}$, which takes the noise perturbation $\bm{Z}$ as input and outputs the least-squares estimate ${\bm{\tau}}$.
To elucidate the dependence on additive noise $\bm{Z}$, we rewrite the optimization in \eqref{eq:varpro} as
\begin{equation*}
\mathop{\mathrm{minimize}}_{\bm{\gamma}\in \mathbb{T}^K} ~ \ell(\bm{\gamma,\bm{Z}})
\end{equation*}
where
\begin{equation}\label{eq:loss_function}
\ell(\bm{\gamma},\bm{Z}) := \norm{\bm{P}_{\bm{\gamma}}^\perp (\bm{G} \bm{\Phi}_{\bm{\tau}} \bm{X} + \bm{Z})}_\mathrm{F}^2.
\end{equation}
Then there exists a nonlinear mapping $\bm{\psi}: \mathbb{C}^{N \times L} \rightarrow \mathbb{T}^K$ given by
\begin{equation*}
   \bm{\psi}(\bm{Z}) \in \mathop{\mathrm{argmin}}_{\bm{\gamma}} \ell(\bm{\gamma},\bm{Z}).
\end{equation*}
The inverse map $\bm{\psi}$ is \textit{locally Lipschitz continuous} at $\bm{0} \in \mathbb{C}^{N \times L}$ if there exists a neighborhood $\mathcal{N}$ of $\bm{0}$ and a constant $C > 0$ such that
\begin{equation}
\label{eq:local_Lip}
\frac{\norm{\bm{\psi}(\bm{Z}_1)- \bm{\psi}(\bm{Z}_2)}_2}{\norm{\bm{Z}_1- \bm{Z}_2}_\mathrm{F}} \leq C, \quad \forall \bm{Z}_1, \bm{Z}_2 \in \mathcal{N}: \bm{Z}_1 \neq \bm{Z}_2.
\end{equation}
The local Lipschitiz constant $L_{\bm{\psi},\mathcal{N}}$ denotes the smallest $C > 0$ satisfying \eqref{eq:local_Lip} and is characterized via the spectral norm of the Jacobian matrix of $\bm{\psi}$ as
\begin{equation*}
L_{\bm{\psi},\mathcal{N}} = \sup_{\bm{Z} \in \mathcal{N}} \norm{\nabla_{\iota(\bm{Z})}\bm{\psi}(\bm{Z})}
\end{equation*}
with $\iota: \mathbb{C}^{L \times N} \rightarrow \mathbb{R}^{2LN}$ defined by $\iota(\bm{Z}) = \begin{bmatrix} \mathrm{Re}(\mathrm{vec}(\bm{Z})) \\ \mathrm{Im}(\mathrm{vec}(\bm{Z})) \end{bmatrix}$ where $\mathrm{vec}(\bm{Z}) \in \mathbb{C}^{LN}$ is obtained by stacking all columns of $\bm{Z}$ vertically.
Note that $\iota$ is an isometric bijection.

For an arbitrary $\bm{Z}$, the Jacobian $\nabla_{\iota(\bm{Z})} \bm{\psi}(\bm{Z})$ of the inverse map $\bm{\psi}$ at $\bm{Z}$ has entries given as multivariate polynomials in $\iota(\bm{Z})$ where the order depends on $N$.
Let $\mathcal{N} \subset \mathbb{C}^{L \times N}$ be a Frobenius-norm ball of radius $r > 0$.
Since $\norm{\nabla_{\iota(\bm{Z})} \bm{\psi}(\bm{Z})}$ is a continuous function in $\iota(\bm{Z})$, in the limit of $r \rightarrow 0$, the local Lipschitz constant $L_{\bm{\psi},\mathcal{N}}$ will converge to $\norm{\nabla_{\iota(\bm{Z})} \bm{\psi}(\bm{0})}$.
Therefore, one may deduce that $\norm{\nabla_{\iota(\bm{Z})} \bm{\psi}(\bm{0})}$ describes the asymptotic sensitivity of the estimation process in the small-noise regime.

The following provides a closed-form expression of $\norm{\nabla_{\iota(\bm{Z})} \bm{\psi}(\bm{0})}$.
It was derived by extending pre-existing results on the differentiation of pseudo-inverse and projection
\cite{golub1973differentiation} to the case of complex-valued matrices. The statement can be verified
with appropriate modification on the transpose operators.
Let $\bm{q}$
denote the partial gradient of the loss function $\ell$ in~\eqref{eq:loss_function} with respect to $\bm{\gamma}$, \emph{i.e.} $\bm{q}(\bm{\gamma},\bm{Z}) = \nabla_{\bm{\gamma}} \ell(\bm{\gamma}, \bm{z})$.
Then it has been shown \cite{basu2000stability} that the \textit{implicit function theorem} \cite{rudin1976principles} (also see \cite[Theorem~6]{basu2000stability}) can be utilized to express the Jacobian of the inverse map $\bm{\psi}$ as
\begin{equation}
\label{eq:jacobian}
\nabla_{\iota(\bm{Z})} \bm{\psi}(\bm{Z})
= - \left( \nabla_{\bm{\gamma}} \bm{q}(\bm{\psi}(\bm{Z}),\bm{Z}) \right)^{-1} \left( \nabla_{\iota(\bm{Z})} \bm{q}(\bm{\psi}(\bm{Z}),\bm{Z}) \right).
\end{equation}
 At zero noise, $\bm{\psi(Z)}$ is reduced to ground-truth parameter $ \bm{\tau}$. Hence, the quantities on the right-hand side of \eqref{eq:jacobian} can be explicitly computed as shown in the following lemma.

\begin{lemma}\label{lem:jacobian}
The Jacobian matrices in the right-hand side of \eqref{eq:jacobian} at $\bm{Z} = \bm{0}$ are written as
\begin{subequations}
\label{eq:jacobian_at_zero}
\begin{equation}
\label{eq:jacobian_at_zero_gamma}
\nabla_{\bm{\gamma}} \bm{q}(\bm{\tau},\bm{0})
=  2 \mathrm{Re} \left(
\overline{\bm{X}} \bm{X}^\mathsf{T}
\odot
\bm{\Phi}_{\bm{\tau}}^* \bm{\Lambda}^* \bm{G}^{*} \bm{P}_{\bm{\tau}}^\perp  \bm{G} \bm{\Lambda} \bm{\Phi}_{\bm{\tau}}  \right)
\end{equation}
and
\begin{equation}
\label{eq:jacobian_at_zero_z}
\nabla_{\iota(\bm{Z})} \bm{q}(\bm{\tau},\bm{0}) = 2
\begin{bmatrix}
\mathrm{Re}
\left( \bm{X}^\mathsf{T} \ast \bm{P}_{\bm{\tau}}^\perp \bm{G} \bm{\Lambda} \bm{\Phi}_{\bm{\tau}} \right) \\
\mathrm{Im}
\left( \bm{X}^\mathsf{T} \ast \bm{P}_{\bm{\tau}}^\perp \bm{G} \bm{\Lambda} \bm{\Phi}_{\bm{\tau}} \right)
\end{bmatrix}^\mathsf{T}
\end{equation}
\end{subequations}
where $\odot$ and $\ast$ respectively denote the Hadamard and Khatri-Rao products, and $\bm{\Lambda} \in \mathbb{R}^{N \times N}$ is a diagonal matrix satisfying $[\bm{\Lambda}]_{i,i} = - \mathsf{j} 2 \pi f_i$ for $i \in [N]$.
\end{lemma}
In the zero-noise case, a closed-form expression for Jacobian norm
$\norm{\nabla_{\iota(\bm{Z})} \bm{\psi}(\bm{0})}$, which characterizes the model sensitivity can be obtained by plugging in \eqref{eq:jacobian_at_zero_gamma}, \eqref{eq:jacobian_at_zero_z} to \eqref{eq:jacobian}. However, it does not explicitly explain how the sensitivity depends on key attributes like minimal separation and PSF characteristics. Therefore, we next present our main result to derive an upper estimate of $\norm{\nabla_{\iota(\bm{Z})} \bm{\psi}(\bm{0})}$ in an interpretable form.
To state the main theorem, we introduce relevant notation.
Let $E_0$ and $E_1$ be the bandlimited energy of the PSF and its derivative in the bandwidth $J_N:= [-\frac{N-1}{2}, \frac{N-1}{2}]$ so that
\begin{align}
\label{eq:E_def}
E_0 &:= \norm{ \widehat{g} \,  \mathbbm{1}_{J_N}}_{L_2}^2, \quad E_1 := \norm{\widehat{g'} \,  \mathbbm{1}_{J_N}}_{L_2}^2,
\end{align}
where $\mathbbm{1}_{J_N}$ is the indicator function of the interval $J_N $ and $\widehat{g'}$ denotes the Fourier transform of the first-order derivative of $g$. Given $\mathcal{C}_0$ the space of continuous functions of the real variable, the total variation norm $\norm{\cdot}_{\mathrm{TV}}$ of a measure $q$ is defined as
\(
    \norm{q}_{\mathrm{TV}} = \displaystyle \sup_{ \substack{h \in \mathcal{C}_0 \\ \norm{h}_{L_\infty} \leq 1}} \int_{-\infty}^{+\infty} h(f) \mathrm{d} q(f).
\)
In the sequel define $\rho$ as the maximum of the normalized total variations of $g$ and $g'$ in the bandwidth $J_N$, \emph{i.e.}
\begin{equation}\label{eq:rho_def}
    \rho :=  \max\left\{ \frac{\norm{|\widehat{g}|^2 \, \mathbbm{1}_{J_N}}_{\mathrm{TV}}}{E_0}, ~ \frac{\norm{|\widehat{g'}|^2 \, \mathbbm{1}_{J_N}}_{\mathrm{TV}}}{E_1} \right\}.
\end{equation}
Intuitively, $\rho$ measures the ``flatness'' of the power spectral densities of $g$ and $g'$ within the interval $J_N$. It decreases as the PSF $g$ and its derivative $g^\prime$ gets narrower in the time domain.

\begin{theorem}\label{THM:MAIN}
Suppose that $\min_{k\neq k'} \inf_{j\in \mathbb{Z}} |\tau_k-\tau_k'+j| \geq \Delta$.
Let $\kappa := \frac{\norm{\bm{X} \bm{X}^*}_{\infty \rightarrow 2}}{\min(\diag(\bm{X} \bm{X}^*))} \geq 1$ be the dynamic range of the spike amplitudes.
If $\Delta > \frac{2}{3} \rho \kappa$, then
\begin{equation}
\label{eq:ub_snorm_jac_zero}
\norm{\nabla_{\iota(\bm{Z})} \bm{\psi}(\bm{0})}
\leq
\frac{\norm{\bm{X}} \sqrt{1 + \frac{2}{3} \rho\Delta^{-1}}}{\sqrt{E_1} \min (\diag(\bm{X} \bm{X}^*)) \left( 1 - \frac{2}{3} \rho  \kappa \Delta^{-1}\right)}.
\end{equation}
\end{theorem}

The parameter $\Delta$ represents the \emph{minimal separation} between any two pulse locations in the support; it has been known since the work of Rayleigh on diffraction to drive the harness of estimating $\bm{\tau}$~\cite{lindberg_mathematical_2012}. Additionally, when the spike amplitudes are drawn according to a statistical model, the dynamic range parameter $\kappa$ is determined by the empirical covariance $\bm{X} \bm{X}^*$ of the spike amplitudes. In the asymptotic number of snapshots, $\bm{X} \bm{X}^*$ converges to the true covariance.
When the amplitudes are uncorrelated, $\kappa$ reduces to the dynamic range of variances of the spike amplitudes. In the other extreme case of a single snapshot, $\kappa$ will be larger with nonzero off-diagonal entries in $\bm{X} \bm{X}^*$.
In this perspective, $\kappa$ also explains how multiple snapshots contribute to mitigating the sensitivity of the parameter estimation.

Note that the right-hand side of \eqref{eq:ub_snorm_jac_zero} is inversely proportional to the scaling of $g$ and $\bm{X}$.
Next, we present a corollary of Theorem~\ref{THM:MAIN} and the following lemma, providing an appropriate normalization of the estimation error by the observation energy.

\begin{lemma}[{\cite[Theorem~1]{ferreira2023conditionNumber}}] For any $\Delta > 0$, we have
\begin{equation*}
        \sigma_{\max} \left(\bm{G} \bm{\Phi}_{\bm{\tau}} \right) \leq  \sqrt{E_0 \left(1 + \frac{1}{2} \rho \Delta^{-1} \right)}.
\end{equation*}
\end{lemma}

To state our result, let us introduce $\mathrm{SNR}:= \frac{\norm{\bm{G} \bm{\Phi}_{\bm{\tau}} \bm{X}}_\mathrm{F}^2}{\norm{\bm{Z}}_\mathrm{F}^2}$.
Then in the limit of $\mathrm{SNR} \rightarrow \infty $, we have
\begin{equation}
\label{eq:noise_prop_snr}
\norm{\bm{\psi}(\bm{Z})-\bm{\tau}}_2
\leq
\norm{\nabla_{\iota(\bm{Z})} \bm{\psi}(\bm{0})} \cdot
\norm{\bm{G} \bm{\Phi}_{\bm{\tau}} \bm{X}}_\mathrm{F}
\cdot
\mathrm{SNR}^{-1/2}.
\end{equation}
The next corollary presents a non-asymptotic upper bound on the noise propagation factor in \eqref{eq:noise_prop_snr}.

\begin{corollary}
\label{cor:noise_propagation}
Suppose that the hypothesis of Theorem~\ref{THM:MAIN} holds.
Then
\begin{equation}\label{eq:noise_amplification}
\norm{\nabla_{\iota(\bm{Z})} \bm{\psi}(\bm{0})} \cdot
\norm{\bm{G} \bm{\Phi}_{\bm{\tau}} \bm{X}}_\mathrm{F}
\leq
\frac{ \eta \, \sqrt{\left(1 + \frac{2}{3} \rho\Delta^{-1}\right)\left(1 + \frac{1}{2} \rho \Delta^{-1}\right)}}{\sqrt{\frac{E_1}{E_0}} \,  \left( 1 - \frac{2}{3} \rho  \kappa \Delta^{-1}\right)}
\end{equation}
where
$\eta:=\frac{\norm{\bm{X}} \cdot \norm{\bm{X}}_\mathrm{F}}{\min (\diag(\bm{X} \bm{X}^*))}$.
\end{corollary}

Corollary~\ref{cor:noise_propagation} describes how the adversarial noise propagates to the estimate by the inverse map $\bm{\psi}$. The parameter $\eta$ is another measure of the dynamic range of the spike amplitudes.
Similar to $\kappa$, the asymptotic number of snapshots is determined by the true covariance matrix. When the amplitudes are uncorrelated, $\eta$ is proportional to $\sqrt{K}$.

The result also suggests improved stability of the inverse map to small-noise for small values of $\rho$. Provided sufficient flatness of the power spectrum densities of $g$ and $g'$, one would expect $\rho$ to be inversely proportional to the bandwidth $|J_N| = N$, \emph{i.e.} $\rho = \mathcal{O}(N^{-1})$.
For example, $\rho \simeq 8/N$ when $g$ is a Dirac function.

\section{Proof of the Main Results}
\label{sec:proof}

\subsection{Proof of Theorem~\ref{THM:MAIN}}

Using~\eqref{eq:jacobian} and by the properties of the spectral norm, we have
\begin{equation}
\label{eq:jacobian_bnd}
\norm{\nabla_{\iota(\bm{z})} \bm{\psi}(\bm{0})}
\leq \frac{\norm{\nabla_{\iota(\bm{Z})} \bm{q}(\bm{\tau},\bm{0})}}{\sigma_{\min}\left(\nabla_{\bm{\gamma}} \bm{q}(\bm{\tau},\bm{0})\right)}.
\end{equation}
To simplify the notation, we let in the sequel $\bm{S}$ the positive definite Hermitian matrix
$\bm{S} = \bm{\Phi}_{\bm{\tau}}^* \bm{\Lambda}^{*} \bm{G}^{*} \bm{P}_{\bm{\tau}}^\perp  \bm{G} \bm{\Lambda} \bm{\Phi}_{\bm{\tau}}$. We continue the proof by controlling the denominator and the numerator on the right-hand side of~\eqref{eq:jacobian_bnd} separately in terms of the quantity $\norm{\bm{S} - E_1 \bm{I}_K}$ from their expression provided by Lemma~\ref{lem:jacobian}.

To provide a lower bound on the denominator on the right-hand side of~\eqref{eq:jacobian_at_zero_gamma}, we write
\begin{align*}
\frac{1}{2} \nabla_{\bm{\gamma}} \bm{q}(\bm{\tau},\bm{0})
&= \mathrm{Re}\left( \overline{\bm{X}} \bm{X}^\mathsf{T} \odot \bm{S} \right) \nonumber \\
&= \mathrm{Re}\left( \overline{\bm{X}} \bm{X}^\mathsf{T} \odot E_1 \bm{I}_K + \overline{\bm{X}} \bm{X}^\mathsf{T} \odot (\bm{S} - E_1 \bm{I}_K) \right) \nonumber \\
&= E_1 \overline{\bm{X}} \bm{X}^\mathsf{T} \odot \bm{I}_K + \mathrm{Re} \left( \overline{\bm{X}} \bm{X}^\mathsf{T} \odot (\bm{S} - E_1 \bm{I}_K) \right),
\end{align*}
where the last identity holds since $\overline{\bm{X}} \bm{X}^\mathsf{T} \odot \bm{I}_K$ is a diagonal matrix with nonnegative entries.
By~\cite[Theorem~1]{zhan1997inequalities}, we have
\begin{align*}
\norm{\mathrm{Re} \left( \overline{\bm{X}} \bm{X}^\mathsf{T} \odot (\bm{S} - E_1 \bm{I}_K) \right)}
& \leq \norm{ \bm{\overline{\bm{X}} \bm{X}^\mathsf{T}} \odot (\bm{S} - E_1 \bm{I}_K) } \nonumber \\
& \leq \norm{\bm{\overline{\bm{X}} \bm{X}^\mathsf{T}}}_{\infty \rightarrow 2} \norm{\bm{S} - E_1 \bm{I}_K}.
\end{align*}
By substituting the expression of $\bm{A}$ and $\bm{S}$, we obtain
\begin{multline}\label{eq:grad_gamma}
\frac{1}{2} \sigma_{\min}(\nabla_{\bm{\gamma}} \bm{q}(\bm{\tau},\bm{0}))
\geq
E_1 \min(\diag(\bm{X} \bm{X}^*))  \\
 - \norm{\bm{S} - E_1 \bm{I}_K} \cdot \norm{\bm{X} \bm{X}^*}_{\infty \rightarrow 2}.
\end{multline}

On the other hand, the numerator on the right-hand side of~\eqref{eq:jacobian_bnd} can be upper-bounded with the inequality $\norm{\bm{A} \ast \bm{B}} \leq \norm{\bm{A}} \cdot \norm{\bm{B}}$ for any matrices $\bm{A}$ and $\bm{B}$ of the same number of columns. This yields
\begin{align}\label{eq:grad_iZ}
& \frac{1}{2} \norm{\nabla_{\iota(\bm{Z})} \bm{q}(\bm{\tau},\bm{0})}
\leq \norm{\bm{X}} \cdot \norm{\bm{P}_{\bm{\tau}}^\perp \bm{G \Lambda} \bm{\Phi}_{\bm{\tau}}} \nonumber \\
& \quad \leq \norm{\bm{X}} \cdot {\left\Vert \bm{S} \right\Vert}_2^{1/2}
\leq \norm{\bm{X}} \cdot \sqrt{E_1 + {\norm{\bm{S} - E_1 \bm{I}_K}}}.
\end{align}

Lemma~\ref{lem:bound_S} proposes a bound on the quantity $\norm{\bm{S} - E_1 \bm{I}_K}$ in terms of the separation between the pulses $\Delta$ and the PSF-related parameters $E_1$ and $\rho$ defined in Equations~\eqref{eq:E_def} and~\eqref{eq:rho_def}.
These bounds are derived via a generalization of the Beurling--Selberg approximation by Vaaler~\cite{vaaler1985ExtremalFunctions}. Its proof is deferred to Section~\ref{subsec:lemma_beurling} for readability.

\begin{lemma}\label{lem:bound_S}
\label{lem:exsv_schur_comp}
Assume that $g(t) \in L_2(\mathbb{R})$ is derivable with $g^{\prime}(t) \in L_2(\mathbb{R})$ then for $\Delta > \frac{2}{3} \rho$, we have
\begin{equation}\label{eq:bound_quotient}
    \norm{\bm{S} - E_1\bm{I}_K}
    \leq \frac{2}{3} E_1 \rho  \Delta^{-1}.
\end{equation}

\end{lemma}

One immediately concludes on the desired statement with Lemma~\ref{lem:bound_S}, and by substituting Equation~\eqref{eq:bound_quotient} into Equations~\eqref{eq:grad_gamma} and~\eqref{eq:grad_iZ}. \hfill \IEEEQEDhere

\subsection{Proof of Lemma~\ref{lem:bound_S}}\label{subsec:lemma_beurling}

Let $\alpha = \sqrt{\frac{E_1}{E_0}}$ be a scaling factor, define $\bm{U} = [\alpha \bm{G} \bm{\Phi}_{\bm{\tau}}, \; \bm{\Lambda} \bm{G} \bm{\Phi}_{\bm{\tau}}]$, and $\bm{M} = \bm{U}^{\ast} \bm{U}$.
By a direct calculation, we have the block decomposition
\begin{align*}
\bm{M} = \begin{bmatrix}
\alpha^2 \bm{\Phi}_{\bm{\tau}}^{\ast} \bm{G}^*  \bm{G} \bm{\Phi}_{\bm{\tau}}
&
\alpha \bm{\Phi}_{\bm{\tau}}^{\ast} \bm{G}^* \bm{\Lambda} \bm{G} \bm{\Phi}_{\bm{\tau}}
\\
\alpha \bm{\Phi}_{\bm{\tau}}^{\ast} \bm{G}^* \bm{\Lambda}^* \bm{G} \bm{\Phi}_{\bm{\tau}}
& \bm{\Phi}_{\bm{\tau}}^{\ast} \bm{G}^* \bm{\Lambda}^* \bm{\Lambda} \bm{G} \bm{\Phi}_{\bm{\tau}}
\end{bmatrix}.
\end{align*}
By the linear independence of trigonometric polynomials and their derivatives, and since $\bm{G}$ has at least $2K$ non-zero diagonal entries, the matrices $\bm{G} \bm{\Phi}_{\bm{\tau}}$, $\bm{\Lambda}\bm{G} \bm{\Phi}_{\bm{\tau}}$, and  $\bm{U}$ are full column rank. This implies the matrix $\bm{M}$, and its two diagonal blocks are invertible. Hence, from the Schur block inversion formula, we have
\begin{align*}
    \bm{M}^{-1} = \begin{bmatrix}
        \ast & \ast \\
        \ast & \bm{S}^{-1}
    \end{bmatrix},
\end{align*}
where we neglected the derivation of blocks marked with an asterisk. Since $\bm{M}$, $\bm{S}$ and their inverse are positive-definite, one may write
\begin{equation*}
    \lambda_{\max} \left(\bm{M}^{-1} \right) \geq  \lambda_{\max} \left( \bm{S}^{-1} \right), \quad \lambda_{\min} \left(\bm{M}^{-1} \right) \leq  \lambda_{\min} \left( \bm{S}^{-1} \right).
\end{equation*}
It comes with $\lambda_{\min}(\bm{Q}^{-1}) = \lambda_{\max}(\bm{Q})^{-1}$, and $\lambda_{\max}(\bm{Q}^{-1}) = \lambda_{\min}(\bm{Q})^{-1}$ for any positive definite matrix $\bm{Q}$ on the inequalities
\begin{align}\label{eq:bound_S_M}
    \lambda_{\max} \left(\bm{M} \right) & \geq  \lambda_{\max} \left( \bm{S} \right),  & \lambda_{\min} \left(\bm{M} \right) & \leq  \lambda_{\min} \left( \bm{S} \right).
\end{align}
With~\eqref{eq:bound_S_M}, we seek to bound the extremal eigenvalues of $\bm{M}$. Relevant bounds are provided in the following Lemma, recalled from~\cite{ferreira2023conditionNumber} eigenvalues of $\bm{M}$, and rely on the Beurling--Selberg extremal approximation of functions with bounded variation.
\begin{lemma}\label{lem:second_order} For any $\Delta > \frac{2}{3} \rho \kappa$, one has the inequalities
\begin{align*}
        \lambda_{\min} \left(\bm{M} \right) & \geq E_1(1-\rho\Delta^{-1}), & \lambda_{\max} \left(\bm{M} \right) & \geq E_1(1+\rho\Delta^{-1}).
\end{align*}
\end{lemma}
One concludes immediately with~\eqref{eq:bound_S_M} and Lemma~\ref{lem:second_order}. \hfill \IEEEQEDhere

\section{Numerical Illustration}
\label{sec:numerical}

In this section, we numerically validate the effectiveness of our sensitivity analysis for the case of Dirac and Gaussian PSFs. The Gaussian PSF is given by
$g(t) = \frac{1}{\sqrt{2\pi}\sigma} \exp\left(-\frac{t^2}{2\sigma^2}\right)$,
where the parameter $\sigma > 0$ determines the effective width of the support.
In this numerical illustration, we set the number of spikes to $K=3$ and the number of Fourier measurements to $N = 501$.
The spike locations are generated randomly so that the minimum separation is no smaller than the parameter $\Delta$.
The observation is maximized over $50$ realizations in the Monte Carlo to simulate the worst-case scenario.
The spike amplitudes are chosen at random and normalized to satisfy $\frac{1}{L} \bm{X} \bm{X}^* = \bm{I}_K$, which corresponds, provided the spike amplitudes are uncorrelated, to the asymptotic case of an infinity of snapshots goes to infinity.

\begin{figure}[t]
    \centering
    \begin{tabular}{cc}
    \includegraphics[trim={0.15in 0in 0.15in 0in},clip,width=0.45\columnwidth]{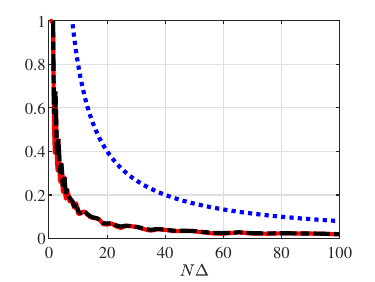} & \includegraphics[trim={0.15in 0in 0.15in 0in},clip,width=0.45\columnwidth]{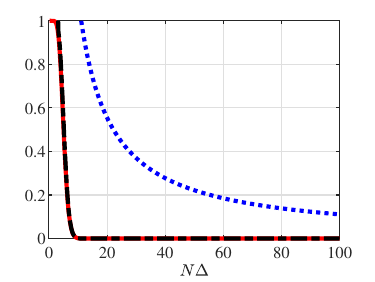}
    \\
    {\footnotesize (a) Dirac PSF} & {\footnotesize (b) Gaussian PSF ($\sigma = 0.02$)}
    \end{tabular}
    \caption{Empirical realization of the quantities $\norm{\bm{S}-E_1\bm{I}_K}$ (in red/solid), $\norm{\bm{M}-E_1\bm{I}_K}$ (in black/dash-dot), and of theoretical upper bound Equation~\eqref{eq:bound_quotient}  from Lemma~\ref{lem:exsv_schur_comp} (in blue/dotted), for varying $\Delta$.
    Herein, $N=501$.}
    \label{fig:lemma_plots}
\end{figure}

\begin{figure}[t]
    \centering
    \begin{tabular}{cc}
	\includegraphics[trim={0.1in 0in 0.15in 0in},clip,width=0.45\columnwidth]{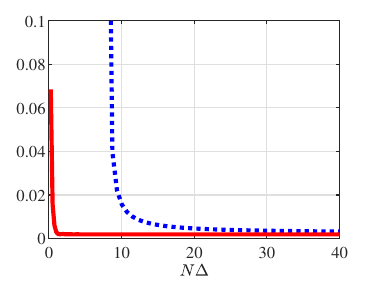} & \includegraphics[trim={0.1in 0in 0.15in 0in},clip,width=0.45\columnwidth]{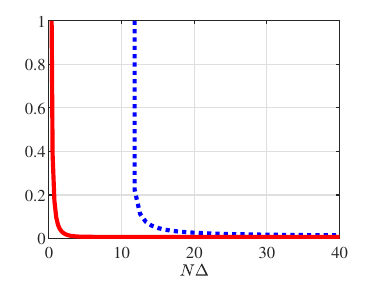}
    \\
    {\footnotesize (a) Dirac PSF} & {\footnotesize (b) Gaussian PSF ($\sigma = 0.02$)}
    \end{tabular}
    \caption{Empirical realization of the noise propagation factor $\norm{\nabla_{\iota(\bm{Z})} \bm{\psi}(\bm{0})} \cdot
\norm{\bm{G} \bm{\Phi}_{\bm{\tau}} \bm{X}}_\mathrm{F}$ (in red/solid), and the upper bound~\eqref{eq:noise_amplification} in  Corollary~\ref{cor:noise_propagation} (in blue/dotted) per varying $\Delta$.}
    \label{fig:theorem_plots}
\end{figure}

Figure~\ref{fig:lemma_plots} compares empirical realization of the quantities $\norm{\bm{S}-E_1\bm{I}_K}$ and $\norm{\bm{M}-E_1\bm{I}_K}$ with their theoretical upper bound given by Equation~\eqref{eq:bound_quotient} in Lemma~\ref{lem:exsv_schur_comp}.
For both PSFs, the spectral distance between $\bm{S}$ and the scaled identity $E_1 \bm{I}_K$ improves with the time-bandwidth product $N\Delta$, suggesting better conditioning. This observation corroborates with the Lemma~\ref{lem:exsv_schur_comp}, which provides a valid upper bound for large enough values of $N\Delta$, and exhibits similar trends in the asymptotic.
We also note that the empirical conditioning of the matrix $\bm{S}$ and the established theoretical bound are higher for the Gaussian PSF than for the Dirac, the latter being a concentrated pulse with a lower spectral flatness factor $\rho$.

\begin{figure}[t]
\centerline{\includegraphics[trim={0.05in 0in 0.15in 0in},clip,width = 0.5\linewidth]{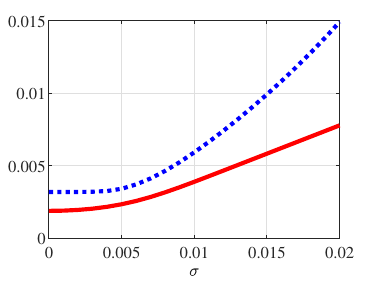}}
\caption{Illustration of the noise propagation factor (in red/solid) and its upper bound in Corollary~\ref{cor:noise_propagation} (in blue/dotted) for the Gaussian PSF per varying $\sigma$ ($N = 501$, $\Delta = 0.25$, $N\Delta \approx 125$).}
\label{fig:varying_sigma}
\end{figure}

Next, we validate the statement of~Corollary~\ref{cor:noise_propagation} throughout numerical simulations.
Figure~\ref{fig:theorem_plots} compares the noise propagation factor and its upper bound for both Dirac and Gaussian PSF. The empirical realization of the noise propagation factor (NPF) steeply increases when $N \Delta$ is smaller than a threshold around $1$, which corresponds to the Rayleigh resolution limit, indicating a highly sensitive regime.
On the other hand, the NPF converges to a constant value as $N \Delta$ increases, landing on the stable estimation regime. The dependence on the factor $N\Delta$ for the stability of pulse-localization has been widely studied for stochastic noise. This experiment illustrates
a similar phenomenon for adversarial noise.
Meanwhile, the bound by Corollary~\ref{cor:noise_propagation} provides a valid upper bound for sufficiently large $N\Delta$. It is also larger than the target quantity by a constant factor.
Besides this conservativeness, the upper bound reflects the overall trend of NPF and is useful since it provides a simple interpretable relation to minimum separation, PSF characteristics, and spike amplitudes.
Figure~\ref{fig:varying_sigma} pictures the NPF and its theoretical upper bound as a function of the width $\sigma$ of a Gaussian PSF. It confirms the intuition that wider convolution kernels are more sensitive to adversarial noise.

\section{Conclusion}
\label{conclsion}
This paper presents a fundamental sensitivity analysis of the spike localization problem in the presence of adversarial noise and arbitrary PSF. The analysis is based on the local Lipschitz property of the inverse map that controls the effect of the measurement noise on parameter estimation error and quantifies the high and low noise sensitivity regions.
We focused on the high-SNR scenario where the local Lipschitz converges to the spectral norm of the noiseless Jacobian of the inverse map.
This setting serves as a baseline measure of stability.
Our main result derives an interpretable upper bound on this quantity in terms of the minimum separation, the shape of the PSF, and the spike amplitude distribution.
The numerical section bolsters the efficacy of the derived bounds.

\clearpage

\renewcommand*{\bibfont}{\footnotesize}
\printbibliography

\end{document}